\documentclass[12pt,authoryear]{article}
\usepackage{mathbbold}
\usepackage{amsfonts}
\usepackage{mathrsfs}
\usepackage{amsthm}
\usepackage{multirow}
\usepackage{bbding}
\usepackage{amssymb}
\usepackage{amsmath}
\usepackage{graphicx,color}
\usepackage[rm,center,compact]{titlesec}
\usepackage[authoryear,round,]{natbib}
\newcommand\bbf[1]{\mathchoice{\hbox{\boldmath$\displaystyle#1$}}
{\hbox{\boldmath$\textstyle#1$}} {\hbox{\boldmath$\scriptstyle#1$}}
{\hbox{\boldmath$\scriptscriptstyle#1$}} }
\newcommand{\bea}{\begin{eqnarray}}
\newcommand{\eea}{\end{eqnarray}}
\newcommand{\Bea}{\begin{eqnarray*}}
\newcommand{\Eea}{\end{eqnarray*}}
\newcommand{\ba}{\begin{array}}
\newcommand{\ea}{\end{array}}
\newcommand{\bt}{\begin{tabular}}
\newcommand{\et}{\end{tabular}}
\newcommand{\btb}{\begin{table}}
\newcommand{\etb}{\end{table}}
\newcommand{\bc}{\begin{center}}
\newcommand{\ec}{\end{center}}
\newcommand{\beq}{\begin{equation}}
\newcommand{\eeq}{\end{equation}}

\makeatletter

\newcommand{\Rmnum}[1]{\expandafter\@slowromancap\romannumeral #1@}
\makeatother

\begin{document}
\bibliographystyle{plain}
\title{Asymptotic Composite Estimation}
\author{
Lu Lin$^1$,  Feng Li$^2$, Kangning Wang$^1$ and Lixing Zhu$^3$\footnote{Lu Lin was supported by NNSF project (11171188 and 11231005) of China, Mathematical
Finance-Backward Stochastic Analysis and Computations in Financial
Risk Control of China (11221061), NSF and SRRF projects (ZR2010AZ001
and BS2011SF006) of Shandong Province of China. Lixing Zhu
was supported by a grant from the University Grants Council of Hong
Kong, Hong Kong, China.}
\\ $^1$Shandong University,  Jinan, China
\\$^2$Zhengzhou Institute of Aeronautical Industry Management, China
\\$^3$Hong Kong Baptist University, Hong Kong}
\date{}
\maketitle

\vspace{-4ex}

\begin{abstract} \baselineskip=16pt
Composition methodologies in the current literature are mainly to promote estimation efficiency via direct composition,  either,  of initial estimators or of objective functions. In this paper, composite estimation is investigated for both estimation efficiency and bias reduction. To this end, a novel method is proposed  by utilizing a  regression relationship between initial estimators and values of model-independent parameter in an asymptotic sense. The resulting estimators could have smaller limiting variances than those of initial estimators, and for nonparametric regression estimation, could also have faster convergence rate than the classical optimal rate that the
corresponding  initial estimators can achieve.  The simulations are carried out to examine its performance in finite sample situations.




\end{abstract}

\newpage
\baselineskip=21pt

\newpage

\setcounter{equation}{0} \titlelabel{\thetitle.\quad}
\section{INTRODUCTION}

Composition methodologies in statistics have received much attention
in the literature. The earlier work may ascend to jackknife
(\citealt{Quenouille1}, \citealt{Quenouille2}, \citealt{Gray-Sch},
\citealt{Tuky}), a special composition approach that combines
leave-one-out versions (or leave-many-out versions) of a traditional
estimator (e.g., the least squares estimator) to construct an
improved estimator. For a comprehensive review see \cite{Miller}.
Recently the notion of composition has been further developed to
several settings mainly for enhancing estimation efficiency.
\citet{Zou-Y} proposed a composite quantile linear regression via
directly combining objective functions, by which the estimation
efficiency is improved. \citet{Kai-L-Z0} extended it to construct
efficiency-improved nonparametric regression estimation through
directly combining the initial estimators. For the further
developments of this methodology in
semiparametric settings, see \citet{Kai-L-Z1}. 
Composite models such as
model averaging are obtained in spirit from the composition idea. By
averaging the selected models beforehand, a refined model can be
obtained; see for example \citet{Wang-Z-Z}, \citet{Hansen}
and \citet{Hoeting}.

From all the aforementioned works, although they respectively treat
their related models for composite estimation construction, we  note
that, to construct a composite estimator, a model-independent
parameter plays a crucial role. This parameter is not the one of
interest for us to estimate, but with different values, several
initial estimators for the parameter of interest can be defined, and then a composite estimator
can be constructed. This is the common feature in all composite
methodologies in the literature. The
examples of  model-independent parameter are the size of blocks in composite
likelihood, the quantile in quantile regression estimation, and the bandwidth in kernel estimation for
nonparametric regression.

It is worthwhile to note the following issues that are of interest to answer. Most of the current composition methodologies in the literature have been developed from case to case. It is of interest to develop a generic framework for composition methodology. To this end, the key is to establish a generic relationship between
estimation and model-independent parameter such that it can be used as a basis for  composition estimation construction. Two of the popularly used
approaches  in the literature have the potential. First is the use of composite objective function. An example
is \citet{Zou-Y} who proposed composite quantile regression (CQR)
with improved estimation efficiency in parametric setup. But \citet{Sun-L} showed that for nonparametric quantile
regression, the weights in composite objective function
asymptotically play no role in enhancing estimation efficiency. The other is to directly combine  initial estimators to form a composite estimator. This method usually cannot however work on bias reduction
when initial estimators are biased such as nonparametric regression
estimation. It is worth pointing out that bias reduction is another important issue as most of existing  methods can only provide biased estimations.

In contrast, we find  that the asymptotic
representations of several estimations can offer us a way to
establish a general framework: the asymptotic composite regression
(ACR). This method has the following desirable features.

\begin{enumerate}
\item ({\it Generality}) The generic framework allows that, as long as
an estimator has an asymptotically linear representation with a
model-independent parameter, a composite estimator can then be constructed by
a regression combination of several initial estimators
according to different values of this parameter.
\item
({\it Variance reduction}) By selecting proper weights, the ACR is shown to be asymptotically more efficient than those obtained by existing composite methods such as
the composite maximum likelihood and the composite least squares.
\item ({\it Bias reduction}) This is particularly useful for bias estimation that is usually the case in the literature. 
This advantage of the ACR could result
in faster convergence rate of biased estimation. For example, under the same regularity conditions, the corresponding ACR of the biased Nadaraya-Watson estimator of
nonparametric regression
can have faster convergence rate than {\it the classical
optimal one}. It is worthwhile to point out that although the ACR estimator seems still to have a kernel estimation type, the above rate-accelerated property is acquired by composition, rather than by a delicately chosen kernel function. Thus, the Nadaraya-Watson estimator cannot possess this property. Further, the composition may be readily applied to other nonparametric smoothing estimations.\end{enumerate}

The rest of the paper is organized as follows. In Section~2 we review
the asymptotic representation of parametric estimation and further
examine three examples to motivate a general framework of
relationship between
 estimator and model-independent parameter. In Section~3,
the ACR is defined and the relevant parametric and nonparametric
estimations are obtained. In Section~4, the accelerated convergence and efficiency
of the new estimators are investigated, and the applications for the
three important models are presented. Simulation studies are given in Section~5 and the
proofs of the theorems are postponed to the Appendix.

\setcounter{equation}{0}
\titlelabel{\thetitle.\quad}
\section{MOTIVATING EXAMPLES AND ASYMPTOTIC REPRESENTATION}

To motivate the methodology development, we first review
asymptotic representations of parametric and nonparametric estimations in several settings. Let $F$ be the
true distribution function of a random variable $X$ and $F_n$ be the
empirical distribution function based on i.i.d observations
$X_1,\cdots,X_n$ from $X$. Consider functional estimators of a
parameter $\theta=T(F)$ of the form $\hat\theta=T(F_n)$ for some
smooth functional $T$ having the influence function
$$I(x)=\lim_{\varepsilon\rightarrow 0}[T((1-\varepsilon)F+\varepsilon\delta_x)
-T(F)]/\varepsilon,$$ where $\delta_x$ is the unit point mass at
$x$. Under some regularity conditions \citep[see, e.g.,][]{Shao}, we have the
following asymptotic representation:
$$\hat\theta-\theta=\frac{1}{n}\sum_{i=1}^nI(X_i)
+\epsilon_n,$$ where $\epsilon_n=O_p(1/n)$ with a mean of order
$O(1/n)$ and a variance of order $O(1/n^2)$. Particularly, for  the
maximum likelihood estimator, $\theta=T(F)$ is defined as the
solution of the equation $\int(\partial/\partial\theta)\log
f_\theta(X)dF(x)=0$ and so $I(x)=J^{-1}(\partial/\partial\theta)\log
f_\theta(X)$, where
$J=-E[(\partial^2/\partial\theta\partial\theta')\log f_\theta(X)]$
and $f_\theta(x)$ is the density function of $X$.

In the above asymptotic representation,
$\frac{1}{n}\sum_{i=1}^nI(X_i)$ is the leading term and determines the
asymptotic property of the estimator $\hat\theta$. In some
situations,  this term could depend on another
parameter. More precisely, the above asymptotic representation often  has
the following form:
\begin{equation}\label{eqno(2.1)}\hat\theta_\tau-\theta=\frac{1}{n}\sum_{i=1}^nI(X_i,\tau)
+\epsilon_n(\tau), \end{equation} for some  parameter $\tau$.  In
the asymptotic representation (\ref{eqno(2.1)}), the model parameter of interest
$\theta$ (or the model itself) is unrelated to the additional
parameter $\tau$ whereas the asymptotic representation (or the
estimator) depends on it. Thus in this paper we call $\tau$ the
model-independent parameter. For illustration, we examine the
following motivating examples.

{\it Example 1 (Linear quantile regression)}. The conditional
$100\tau\%$ quantile of $Y|X$ is
$$\beta^TX+b_\tau,$$
where $b_\tau$ is the $100\tau\%$ quantile of $Y-\beta^TX$. 
Without loss of generality, assume that
$E(Y-\beta^TX-b_\tau|X)=0$. The quantile regression estimator of $(b_\tau,\beta^T)^T$
can be obtained as
$$\left(\begin{array}{cc}\hat b_\tau\\\hat\beta_\tau\end{array}\right)=\arg\min_{\hspace{-3ex}b_\tau,\beta}
\sum_{i=1}^n\rho_\tau (Y_i-b_\tau-\beta^TX_i),$$ where
$\rho_\tau(t)=\tau t_++(1-\tau)t_-$ is the so-called check function
with $+$ and $-$ standing for  positive and negative parts,
respectively. Denote $F_i(y)=F(y|X_i)=P(Y_i<y|X_i)$ and suppose that
$F_i(y),i=1,\cdots,m$, are i.i.d. with a common density function
$f(y)>0$ for all $y$. Under some regularity conditions (see, e.g., \citealt{Bahadur,Kiefer,Koenker}),
we have the following
Bahadur representation:
$$\hat\beta_\tau-\beta=\xi(\tau,\beta)\varphi_n +\epsilon_n(\tau),$$ where
$\varphi_n=n^{-1/2}$, $\epsilon_n(\tau)=O_p(n^{-3/4})$,
\begin{eqnarray}\nonumber\xi(\tau,\beta)&=&f^{-1}(Q(\tau))D^{-1}
\frac{1}{\sqrt{n}}\sum\limits_{i=1}^n X_i (\tau-I(Y_i\leq
b_\tau+\beta^T X_i)),\\ \nonumber
D&=&\lim\limits_{n\rightarrow\infty}\frac{1}{n}\sum\limits_{i=1}^n
X_iX_i^T,
\end{eqnarray} and $Q(\tau)=F^{-1}(\tau|X)$, the
$\tau$th quantile of $Y$. Here $\xi(\tau,\beta)$ is of order
$O_p(1)$, and  in the next section we will show that under a mild
condition $\xi(\tau,\beta)$ can be estimated. We can see that the
regression coefficient $\beta$ is independent of $\tau$, but the
asymptotic representation of the estimator $\hat\beta_\tau$ depends on $\tau$.  The argument can be applied to nonlinear parametric models. \hfill$\Box$

{\it Example 2 (Nonparametric regression)}. Consider the following
nonparametric regression:
$$Y=r(X)+e,$$ where $r(x)$ is a smooth nonparametric regression function for
$x\in [0,1]$, the error term satisfies $E(e|X)=0$ and
$Var(e|X)=\sigma^2$. We now give two asymptotic representations for
the kernel estimator of $r(x)$ with $x\in(0,1)$. As is known,
$x\in(0,1)$ is not a necessary constraint, we use it only for simplicity
of presentation. It is well known that under certain regularity conditions with second order continuous and bounded derivatives,
a commonly used kernel estimator $\hat r_\tau(x)$ (e.g., Nadaraya-Watson
estimator, we write it as the N-W estimator throughout the rest of the paper) of the regression function $r(x)$ has  the mean value:
$$E(\hat
r_\tau(x))=r(x)+\frac{1}{2}\Big\{r''(x)+2\frac{r'(x)f_X'(x)}{f_X(x)}\Big\}
\mu_2(K)h^2+O(h^4),\ x\in(0,1),$$ where $f_X(x)$ is the density
function of $X$, $\mu_2(K)=\int u^2K(u)du$, $K(x)$ is a kernel
function and $h$ is a bandwidth satisfying $h=\tau n^{-\eta}$ for
constants $\tau>0$ and $0<\eta<1$. Then we have the following
asymptotic representation
$$\hat r_\tau(x)-r(x)=\xi(\tau)\varphi_{1n}+\epsilon_n(\tau),\ x\in(0,1),$$ where
$\xi(\tau)=\tau^2$,
$\varphi_{1n}=\frac{1}{2}\Big\{r''(x)+2\frac{r'(x)f_X'(x)}{f_X(x)}\Big\}\mu_2(K)n^{-2\eta}$
and $\epsilon_{n}=\hat r_\tau(x)-E(\hat r_\tau(x))+O(n^{-4\eta})$.
Here $\epsilon_n$ has mean of order $O(n^{-4\eta})$ and variance of
order $n^{-(1-\eta)}$ and therefore is of order $o_p(n^{-2\eta})$
provided that $0<\eta<1/5$.

Also we can use the Bahadur representation (see, e.g.,
\citealt{Bhattacharya-G,Chaudhuri,Hong}) to construct a relationship
between the estimator and the model-independent parameter. Under
regularity conditions (including the condition in Theorem 3.4(2)
given in Section 3), the N-W estimator $\hat r_\tau(x)$ has
following Bahadur representation:
$$\hat r_\tau(x)-r(x)= \xi(\tau,r)\varphi_{2n}+\epsilon_n(\tau),\ x\in(0,1),$$
where $\epsilon_n(\tau)$ is of order $O_p(n^{-3(1-\eta)/4})$,
$\varphi_{2n}=n^{-(1-\eta)/2}$,
$$\xi(\tau,r)=n^{-(1+\eta)/2}v^{-1}_\tau(x)\sum_{i=1}^nK_\tau(X_i-x)(Y_i-r(x)), \ x\in(0,1),$$
$v_\tau(x)=\int K(u)f_X(x+hu)du$ and $K_\tau(x)=h^{-1}K(x/h)$ with
$h=\tau n^{-\eta}$. Here $\xi(\tau,r)$ is of order $O_p(1)$ and
obviously can be estimated.

The two representations above show that the asymptotic
representations for nonparametric regression are also related to a
model-independent parameter $\tau$ (or $h$). \hfill$\Box$

{\it Example 3 (Blockwise likelihood)}.  Blockwise
composite likelihood (see, e.g., \citealt{Varin-R-F}) is usually used
for  models with dependent data. In this example, we  consider the blockwise
empirical likelihood. Let $Y_1,\cdots, Y_n$ be dependent
observations from an unknown $d$-variate distribution $f(y;\theta)$,
where the parameter vector $\theta\in \Theta\subset R^p$. The
information about $\theta$ and $f(y;\theta)$ is available in the
form of an unbiased estimating function $u(y;\theta)$, i.e.
$E(u(Y;\theta^0))=0$, where $\theta^0$ is the true value of $\theta$
and $u(y; \theta)$ is a given function vector:
$R^d\times\Theta\rightarrow R^r$ with $r\geq p$. Let $M$ and
$L_\tau$ be integers satisfying $M=[n^{1-c}]$ and $L_\tau=[\tau
n^{1-c}]$ for some constants $0<c\leq1$ and $0<\tau\leq1$, where $[x]$
stands for the integer part of $x$. Denote
$B_i=(Y_{(i-1)L_\tau+1},\cdots,Y_{(i-1)L_\tau+M})^{\tau}$,
$i=1,\cdots,Q_\tau$, where $Q_\tau=[(n-M)/L_\tau]+1$. It can be
verified that $Q_\tau = O(n^c)$. We can see that $B_i$ are  blocks of
observations, $M$ is the window-width, and $L_\tau$ is the separation
between the block start points. The observation blocks
$B_i$  are used to construct the following estimating function:
$$U_{i}(\theta,\tau)=\frac{1}{M}\sum_{k=1}^{M} u(Y_{(i-1)L_\tau+k};\theta).$$
Then, the blockwise empirical Euclidean log-likelihood ratio for
dependent data is defined as
$$l_\tau(\theta)=\sup\left\{-\frac{1}{2}\sum_{i=1}^{Q_\tau}(Q_\tau p_i-1)^2\Big|
\sum_{i=1}^{Q_\tau} p_i=1,p_i\geq
0,\sum_{i=1}^{Q_\tau}p_iU_{i}(\theta,\tau)=0\right\},$$ and the empirical
Euclidean likelihood estimator of $\theta$ is defined as
$$\hat\theta_\tau=\sup_{\theta\in\Theta}l_\tau(\theta).$$ Here we only
consider the case of $p=r=1$.
It follows
from the asymptotic representation given in the proof of Theorem 2 of \citet{Lin-Zh} that under certain regularity conditions, the
following asymptotic representation holds:
$$\hat\theta_\tau-\theta=\xi(\tau,\theta)\varphi_n+o_p\Big(\frac{1}{\sqrt{n}}\Big),$$
where $$\xi(\tau,\theta)=\sqrt{n}\,\bar U(\theta,\tau), \
\varphi_n=\frac{1}{\sqrt{n}\,\Delta(\theta)},$$ $\bar
U(\theta,\tau)=\frac{1}{Q_\tau} \sum_{i=1}^{Q_\tau}\,U_{i}(\theta,\tau)$ and
$\Delta(\theta)=E(u'(Y;\theta))$ with
$u'(y;\theta)$ being the derivative of $u(y;\theta)$ with respect to $\theta$. Here
$\xi(\tau,\theta)$ is of order $O_p(1)$ and the model parameter
$\theta$ is also free of $\tau$ but the asymptotic representation
given above depends on it. For this estimator, $c$ could also be regarded as a
model-independent parameter. But for
simplicity, we do not take this case into account.
\hfill$\Box$

A common feature of all the asymptotic representations in Examples
1-3 is the formulation of (\ref{eqno(2.1)}). Also we can easily find
other examples to have the common feature of this formulation. We list a few here: the relationship between
penalty based estimators (e.g., the LASSO estimator) and penalty
parameter;  between B-spline estimator and the
number of knots;  between wavelet estimator and the
bandwidth. Thus, this generic method may readily be extended to handle other estimations with both bias  and variance reduction.

\setcounter{equation}{0}
\titlelabel{\thetitle.\quad}
\section{ASYMPTOTIC COMPOSITE ESTIMATION}

\subsection{ A Regression Modeling via Asymptotic Representation}
 The asymptotic representations in (\ref{eqno(2.1)}) and
Examples 1-3 reveal the relationship between  model parameter of
interest and model-independent parameter. Thus it offers us an
useful way to construct new composite estimation in a general
framework.

Note that we can define estimators according to different values $\tau_k,
k=1,\cdots,m$, of
the model-independent parameter $\tau$. For example, for quantile regression estimation,
$\tau_k, k=1,\cdots,m$, are different quantile positions; for
nonparametric regression, $\tau_k$ are determined by different
$V_k$-fold cross-validations, $k=1,\cdots,m$. Let
$\hat\theta_{\tau_k}$ be the corresponding estimators. We then
regress $\hat\theta_{\tau_k}$ on $\tau_k$ to construct a new
estimator of $\theta$ as the intercept of the following regression
model (or $\theta$ regression model):
\begin{equation}\label{eqno(3.1-new-0)}\hat\theta_{\tau_k}=\theta+g(\tau_k)+
\epsilon_n(\tau_k),\ k=1,\cdots,m.\end{equation} Here $g(\cdot)$ is
an unknown function. For the sake of identifiability, based on
Examples 1-3, we assume (\ref{eqno(3.1-new-0)}) has the following
framework:
\begin{equation}\label{eqno(3.1)}\hat\theta_{\tau_k}=\theta+
\xi(\tau_k,\theta)\varphi_n+\epsilon_n(\tau_k),\
k=1,\cdots,m,\end{equation} where $\xi(\tau,\theta)$ is a known
function of $(\tau,\theta)$ and is of  order $O_p(1)$, and
$\varphi_n$ may be an unknown function with respect to $\theta$, but
is independent of $\tau$. We further assume the following
condition:
\begin{itemize}
\item[(C1)]
$\varphi_n=O_p(n^{-\delta_1})$ and
$\epsilon_n(\tau_k)=O_p(n^{-\delta_2})$, where $\delta_1$ and $\delta_2$
are positive constants satisfying $\delta_1<\delta_2$.
\end{itemize}
The above condition is not restrictive and several estimators, say,
those in Examples 1-3, satisfy it. The condition determines the
convergence rate of every term on the right-hand side of
(\ref{eqno(3.1)}) and thus $\epsilon_n(\tau_k)$  could be regarded as the error
term. We call model
(\ref{eqno(3.1)}) (or (\ref{eqno(3.1-new-0)})) the asymptotic
composite regression (ACR) because it is established by using the
asymptotic representation of  estimator and a regression idea with the estimator as
the response variable $\hat\theta_{\tau}$, and the model-independent parameter as the
covariate $\tau$. Thus a composite
estimator of $\theta$ is just the estimator of the intercept on the
right-hand side of (\ref{eqno(3.1)}).

\subsection{ Estimation} Because
$\xi(\tau,\theta)$ may be related to $\theta$, we first construct an
initial estimator $\hat\theta$ to replace it. Denote
$\hat\xi(\tau)=\xi(\tau,\hat\theta)$. Here the initial estimator
$\hat\theta$ may depends on $\tau$. We consider two different cases separately.

(1) We first consider the case of $\varphi_n$ being an unknown function.
Because $\delta_2>\delta_1$,
$\xi(\tau_k,\theta)\varphi_n$ is the leading term of equation
(\ref{eqno(3.1)}). In this case we ignore $\epsilon_n(\tau_k)$ and then
construct a composite estimator $\tilde\theta$ of $\theta$ as the first component of the following minimizers:
\begin{equation}\label{eqno(3.3)}\left(\begin{array}{cc}\tilde\theta\\\tilde\varphi_n\end{array}\right)=\arg\min_{\hspace{-3ex}\theta,\varphi_n}
\frac{1}{m}\sum_{k=1}^mw_k
(\hat\theta_{\tau_k}-\theta-\hat\xi(\tau_k)\varphi_n)^2,\end{equation} where
$w_k,k=1,\cdots,m$, are weights satisfying
$\sum_{k=1}^mw_k=1$. The
estimator can be expressed as
\begin{equation}\label{eqno(3.4)}\tilde\theta=\sum_{k=1}^mw_{k}
\hat\theta_{\tau_{k}} -\hat\varphi_n\bar{\hat\xi},
\end{equation} where
$\bar{\hat\xi}=\sum_{k=1}^mw_{k}\hat\xi(\tau_k)$ and
$$\hat\varphi_n=\frac{\sum_{k=1}^mw_{k}
\hat\theta_{\tau_{k}}\left(\hat\xi(\tau_k)- \bar{\hat\xi}\right)}
{\sum_{k=1}^mw_{k}\left(\hat\xi(\tau_k)-\bar{\hat\xi}\right)^2}.$$

(2) If $\varphi_n$ is given,  $\theta$ can be simply estimated as
\begin{equation}\label{eqno(3.5)}\tilde\theta
=\sum_{k=1}^mw_{k}\left(\hat\theta_{\tau_{k}}
-\hat\xi(\tau_k)\varphi_n\right).\end{equation}

We call $\tilde\theta$ defined in (\ref{eqno(3.4)}) and (\ref{eqno(3.5)}) the ACR estimator.
The theoretical properties for the estimators will be given in
Section~4.

\subsection{Estimators for the Three Examples}

Now we  construct the corresponding composite estimators for the three
examples mentioned in Section 2.

{\it (a) Asymptotic composite quantile regression estimation.} For
the quantile regression estimation given in Example~1, suppose that
the conditional density function $f_e(\cdot|X)$ of the error $e$ is
given. We choose the initial estimators $\hat
b_{\tau}$ and $\hat\beta_\tau$ respectively of $b_\tau$ and $\beta$ as the quantile regression estimators defined
in Example~1. 
According to (\ref{eqno(3.5)}), the ACR estimator has the form:
\begin{eqnarray}\label{eqno(3.8)}\hspace{-0.5cm}\tilde\beta&=&
\sum_{k=1}^mw_k \left\{\hat\beta_{\tau_k} - \frac{1}{\hat
f(Q(\tau_k))n}\hat D_n^{-1}\sum\limits_{i=1}^n X_i(\tau_k-I(Y_i\leq
\hat b_{\tau_k}+\hat\beta^{\tau}_{\tau_k} X_i))\right\},\end{eqnarray}
where $\hat f(Q(\tau_k))=f_e(\hat b_{\tau_k}|X)$, $\hat D_n =
\frac{1}{n}\sum\limits_{i=1}^n X_iX_i^{\tau}$ and $\tau_k, k=1,\cdots,m$,
are different quantile positions.

{\it  (b) Asymptotic composite nonparametric regression estimation.}
Here we only use the N-W estimator as the initial estimator, which
is defined as
$$\hat r_\tau(x)=\frac{\sum_{i=1}^nY_iK_\tau(X_i-x)}{\sum_{i=1}^nK_\tau(X_i-x)},\ x\in(0,1).$$
The asymptotic representations in Example 2 and the estimators
(\ref{eqno(3.4)}) and (\ref{eqno(3.5)}) result in that two ACR
estimators $\tilde r_i(x)$ of the regression function $r(x)$ for
$x\in(0,1)$ can be defined as
\begin{eqnarray}\label{eqno(3.9)}&&\nonumber\tilde r_1(x)=\sum\limits_{k=1}^mw_k
\hat r_{\tau_k}(x) -\tilde\varphi_{1n}\overline{\tau^2} \ \mbox{ and } \\
&&\tilde r_2(x)=\sum\limits_{k=1}^mw_k \left(\hat
r_{\tau_k}(x)-n^{-(1-\eta)/2}\hat\xi(\tau_k)\right), \ x\in(0,1),
\end{eqnarray} where $\tau_k$ are about the bandwidths $h_k=\tau_kn^{-\eta}$, $k=1,\cdots,m$,
\begin{eqnarray*}\tilde\varphi_{1n}&=&\frac{\sum\limits_{k=1}^mw_k\hat
r_{\tau_k}(x)(\tau_k^2- \overline{\tau^2})}
{\sum\limits_{k=1}^mw_k(\tau_k^2-\overline{\tau^2})^2},\ \ \overline{\tau^2}=\sum\limits_{k=1}^mw_k\tau^2_k,\\\hat\xi(\tau)&=&n^{-1}
v_{\tau}^{-1}(x)\sum\limits_{i=1}^nK_{\tau}(X_i-x)(Y_i-\hat
r_{\tau}(x)).\end{eqnarray*}
In practical use, we may determine $h_k=\tau_kn^{-\eta}$ by different $V_k$-fold
cross-validations.

{\it (c) (Blockwise empirical likelihood estimation)}. Consider blockwise empirical likelihood in Example~3.
The blockwise empirical Euclidean log-likelihood ratio has the following closed representation:
$$l_\tau(\theta)=-\frac{Q_\tau}{2}\bar U^T(\theta,\tau)S^{-1}(\theta,\tau)\bar U(\theta,\tau),$$ where $S(\theta,\tau)=\frac{1}{Q_\tau}\sum_{i=1}^{Q_\tau}(U_i(\theta,\tau)-\bar U(\theta,\tau))(U_i(\theta,\tau)-\bar U(\theta,\tau))^T$; see. e.g., \citet{Lin-Zh}. Given $\tau=\tau_k$, denote by $\hat\theta_{\tau_k}$ the initial estimators of $\theta$ by maximizing the above likelihood function. For simplicity, we here only consider the case with $p=r=1$, i.e., both the parameter $\theta$ and the unbiased estimating function $u(y;\theta)$  are  scalar. It follows from (\ref{eqno(3.4)}) that the composite
estimator can be expressed as
\begin{equation}\label{eqno(block)}\tilde\theta=\sum_{k=1}^mw_{k}
\hat\theta_{\tau_{k}} -\hat\varphi_n\bar{\hat\xi},
\end{equation} where
$\bar{\hat\xi}=\sum_{k=1}^mw_{k}\hat\xi(\tau_k)$, $\hat\xi(\tau_k)=\sqrt{n}\,\bar U(\hat\theta_{\tau_k},\tau_k)$ and
$$\hat\varphi_n=\frac{\sum_{k=1}^mw_{k}
\hat\theta_{\tau_{k}}\left(\hat\xi(\tau_k)- \bar{\hat\xi}\right)}
{\sum_{k=1}^mw_{k}\left(\hat\xi(\tau_k)-\bar{\hat\xi}\right)^2}.$$

\setcounter{equation}{0}
\section{THEORETICAL PROPERTIES AND OPTIMAL WEIGHTS}
 It is
known that if (\ref{eqno(3.1)}) were a true linear regression model,
the least squares estimator would be unbiased with minimum variance
in certain sense. However, this model is only a linear model in form
whose error term has a bias of order $O(n^{-\delta_2})$ in
probability and the main part tends to zero at a certain convergence
rate.

In this section we suppose $\theta$ is a scalar
parameter for simplicity.
When $\xi(\tau)$ is free of $\theta$, we define the regenerated weights by $$\tilde w_k=w_k-\bar{\xi}\frac{w_k(\xi(\tau_k)-\bar{\xi})}
{\sum_{k=1}^mw_k(\xi(\tau_k)-\bar{\xi})^2},$$ which are free of the initial estimators and still satisfy $\sum_{k=1}^m\tilde w_k=1$. We have the following
theorem.

\noindent{\bf Theorem 4.1.} {\it When  $\xi(\tau)$ is
unrelated to $\theta$, then the ACR estimator $\tilde\theta$ satisfies
$$\tilde\theta-\theta=\sum\limits_{k=1}^m\tilde w_k
\epsilon_n(\tau_k),$$ where $\epsilon_n(\tau_k)$ are the error terms of the asymptotic representation defined in (\ref{eqno(3.1)}).}

{\bf Remark 4.1.} Interestingly, from the representation in the theorem and (C1) we can
see that when $\xi(\tau)$ is free of $\theta$, the convergence rate of the ACR estimator is faster than those of the initial estimators.
The first estimator in (\ref{eqno(3.9)}) has this property; see the details in
Theorem~4.4 below. In other words, the ACR estimator can be super-efficient in certain scenarios.  Remark~4.2 (a) given
below will further verify this point of view. Theorem 4.1 also implies that the ACR estimator is bias-reduced.
In the following, we give a result showing that the ACR method can reduce
the variance in finite sample cases.

Define regenerated weights as
$$\tilde w_k=w_k-\bar{\xi}\frac{w_k(\xi(\tau_k)-\bar{\xi})}
{\sum_{k=1}^mw_k(\xi(\tau_k)-\bar{\xi})^2},\ k=1,\cdots,m.$$
Let ${\bf w} =(w_1,\cdots,w_m)^T$, $\tilde {\bf w} =(\tilde w_1,\cdots,\tilde w_m)^T$ and
$\bf 1$ be a $m$-dimensional column vector with all components 1.

\noindent{\bf Theorem 4.2.} {\it When  $\xi(\tau)$ is free of $\theta$, the
variance of the ACR estimator $\tilde\theta$
can be expressed as
\begin{equation}\nonumber
Var(\tilde\theta)=\tilde {\bf w}^T\Sigma_{\hat{\bbf\theta}}\tilde
{\bf w}.\end{equation} Particularly, when the original weight vector $\bf w$ are chosen by the following equation:
$$\tilde {\bf w}=({\bf 1}^T\Sigma^{-1}_{\hat{\bbf\theta}}{\bf 1})^{-1}\Sigma^{-1}_{\hat{\bbf\theta}}{\bf 1},$$
then,
\begin{equation}\nonumber Var(\tilde\theta)
\leq Cov(\hat\theta_{\tau_k}) \ \mbox{ for }
k=1,\cdots,m.\end{equation}
 }

For the theorem, we have the following remark:

{\bf Remark 4.2.} \begin{itemize}
\item[(a)] Denote ${\bf w}^*=(w_1^*,\cdots,w_m^*)^T=({\bf 1}^T\Sigma^{-1}_{\hat{\bbf\theta}}{\bf 1})^{-1}\Sigma^{-1}_{\hat{\bbf\theta}}{\bf 1}$, which is the optimal weight vector that minimizes $Var(\tilde\theta)=\tilde {\bf w}^T\Sigma_{\hat{\bbf\theta}}\tilde
{\bf w}$ subject to $\sum_{k=1}^m\tilde w_k=1$.
Then, the theorem shows that we should choose the original weights $w_k$ by equations:
$$w_k-\bar{\xi}\frac{w_k(\xi(\tau_k)-\bar{\xi})}
{\sum_{k=1}^mw_k(\xi(\tau_k)-\bar{\xi})^2}=w_k^*,\ k=1,\cdots,m.$$ Denote by $w^0_k,k=1,\cdots,m$, the solutions of the above equations. Note that for arbitrary weights $w_k$, $\sum_{k=1}^mw_k(\lambda_k-\bar\lambda)=0$, and the optimal weights $w^*_k,k=1,\cdots,m$, satisfy $\sum_{k=1}^mw^*_k=1$. Thus, for the solutions of the above equations, the regularization condition still holds, formally,
$$\sum_{k=1}^mw_k^0=1,$$ which is a key condition for bias correction.
However, the above $m$ equations
are nonlinear, we cannot get closed forms of the
solutions  of $m$ unknown weights $w_k,k=1,\cdots,m$. Thus numerical
methods are required.
\item[(b)]
The above two theorems ensures that the ACR can
simultaneously reduce bias and variance in certain cases. This sheds
the insights on the potential merits of the ACR.  Of course, the key
condition is that $\xi(\tau)$ is unrelated to $\theta$. The first
estimator in (\ref{eqno(3.9)}) satisfies this condition. However, this
condition is unsatisfied sometimes. For instance, the estimator in
(\ref{eqno(3.8)}), the second estimator in (\ref{eqno(3.9)}) and the estimator in (\ref{eqno(block)}) do
not satisfy this condition. In this situation, only their
asymptotic properties can be obtained. In the following, we will
investigate the asymptotic properties for the important estimators
(\ref{eqno(3.8)})-(\ref{eqno(block)}) regardless of whether this
condition is satisfied or not.\end{itemize}

Consider the composite quantile regression estimator
(\ref{eqno(3.8)}). In addition to those given in Example 1, we need
the following conditions:
\begin{itemize}\item[(C2)]
$\max\limits_{1\leq i\leq n}\|X_i\|\leq cn^{\nu}$ for some constants
$c>0$ and $0\leq\nu< 1/2$.
\item[(C3)] The conditional density function $f_e(u|x)$ of the error $e$
is continuously differentiable.
\end{itemize}
The following theorem states the asymptotic normality of the
estimator.

\noindent{\bf Theorem 4.3.} {\it For the linear regression model in
Example~1, when (C2) and (C3) hold,  the ACR estimator
(\ref{eqno(3.8)}) has the following asymptotic representation:
\begin{eqnarray}\nonumber \tilde \beta-\beta
=D^{-1}\frac{1}{n}\sum\limits_{i=1}^n X_i\sum_{k=1}^mw_k
f^{-1}(Q(\tau_k)) (\tau_k-I(Y_i\leq b_{\tau_k}+\beta^T
X_i))+O_p(n^{-3/4}).\end{eqnarray} Consequently,
$$\sqrt{n}(\tilde \beta-\beta)\stackrel{\cal D}\longrightarrow
N\left(0,{\bf w}^TA_0{\bf w}D^{-1}\right),$$ where ${\bf
w}=(w_1,\cdots,w_m)^T$ and
$$A_0=\left(\frac{\min(\tau_k,\tau_{k'})
(1-\max(\tau_k,\tau_{k'}))}{f(Q(\tau_k))f(Q(\tau_{k'}))}\right)_{k,k'=1}^m.$$}

{\bf Remark 4.3.} Particularly, when
$w_k=f(Q(\tau_k))/\sum_{k=1}^mf(Q(\tau_k))$, the limiting covariance
of the ACR estimator (\ref{eqno(3.8)}) is the same as that of the
composite estimator proposed by Zou and Yuan (2008). Furthermore, we
can obtain the optimal weights in the following way. By minimizing
the limiting variance given in Theorem~4.3, we see that the optimal
weight vector has the form
$${\bf w}^*=\min_{{\bf 1}^T{\bf w}=1}{\bf w}^TA_0{\bf w}.$$  By
Lagrange multipliers, we see that the optimal weight vector has the
following closed representation:
$${\bf w}^*=\left({\bf 1}^TA^{-1}_0{\bf  1}\right)^{-1}A^{-1}_0{\bf 1},$$
and, as a result, the optimal limiting covariance of
$\sqrt{n}(\tilde \beta-\beta)$ is
$${{\bf w}^*}^TA_0{\bf w}^*D^{-1}=\left({\bf 1}^TA^{-1}_0{\bf  1}\right)^{-1}D^{-1}.$$ With this optimal weight,
the resulting estimator is more efficient than the composite estimator of \citet{Zou-Y}, but is the same as in \citet{Koenker-1984}. In (\ref{eqno(3.8)}), when $f(Q(\tau_k))$ are
estimated consistently, we can get a consistent estimator of the
optimal weight vector ${\bf w}^*$. For a different
problem (\citealt{Fan-W}),  the
choice of the optimal weights was discussed, but, the theoretical justification in the scenario under study was not explored before.

We now investigate the asymptotic property of the estimators in
(\ref{eqno(3.9)}) for the nonparametric regression model defined in
Example 2. We consider the following two regularity conditions
respectively:
\begin{itemize}\item[(C4)] Kernel function $K(u)$ is symmetric with
respect to $u=0$, and satisfies $\int K(u)du=1$, $\int u^2
K(u)du<\infty$ and $\int u^2 K^2(u)du<\infty$. Regression function
$r(x)$ defined in Example 2 and density function $f_X(x)$ of $X$
have the second-order continuous and bounded derivatives and
$f_X(x)>0$ for all $x$.
\item[(C5)]  Kernel function $K(u)$ is symmetric with
respect to $u=0$, and satisfies $\int K(u)du=1$, $\int u^4
K(u)du<\infty$ and $\int u^2 K^2(u)du<\infty$. Functions $r(x)$ and
$f_X(x)$ have the fourth-order continuous and bounded derivatives
and $f_X(x)>0$ for all $x$.
\end{itemize}
It is well known that for the N-W estimator, the convergence rate is related to two factors: bandwidth selection and smoothness of the regression function. Generally speaking, the more smooth the regression function is, the faster the rate can achieve when larger bandwidth  and higher order kernel function are used.  We note that condition (C4) is the typical condition for the N-W estimator when only second order derivatives are assumed for the smoothness of the regression function. However, condition (C5) is of interest. It assumes the smoothness with the fourth order derivatives, but does not require the higher order kernel. For the N-W estimator, its convergence rate cannot be accelerated, whereas the ACR estimator can. The following theorem states these.

Denote $s_k({\bf w})=1-\frac{\overline{\tau^2}
(\tau_k^2-\overline{\tau^2})}
{\sum\limits_{k=1}^mw_k(\tau_k^2-\overline{\tau^2})^2}$,
$g_k=w_k-\overline{\tau^2}\frac{w_k (\tau_k^2-\overline{\tau^2})}
{\sum\limits_{k=1}^mw_k(\tau_k^2-\overline{\tau^2})^2}$ and
\begin{eqnarray}\nonumber &&A_1({\bf w})=\left(\frac{
s_k({\bf w})s_j({\bf w})}{\tau_k\tau_j}\int
K\Big(\frac{u}{\tau_k}\Big)K\Big(\frac{u}{\tau_j}\Big)du\right)_{k,j=1}^m,\\
\nonumber &&A_2=\left(\frac{ 1}{\tau_k\tau_j}\int
K\Big(\frac{u}{\tau_k}\Big)K\Big(\frac{u}{\tau_j}\Big)du\right)_{k,j=1}^m.\end{eqnarray}
We have the following results.

{\bf Theorem 4.4.} {\it Suppose $h_k=\tau_kn^{-\eta},k=1,\cdots,m$,
$0<\eta<1$.\\ (1) Under Condition (C4) or (C5), there is an
$c_n(x)=o(n^{-2\eta})$ or $c_n(x)=
n^{-4\eta}c(x)\sum\limits_{k=1}^mg_k \tau_k^4$ accordingly, the ACR
estimator $\tilde r_1(x)$ in (\ref{eqno(3.9)}) achieves the
following asymptotic normality:
\begin{eqnarray*}\sqrt{n^{1-\eta}}\Big(\tilde
r_1(x)-r(x)-c_n(x)\Big)\stackrel{\cal D}\longrightarrow N\Big(0,{\bf
w}^TA_1({\bf w}){\bf w}\frac{\sigma^2}{f_X(x)}\Big), \
x\in(0,1).\end{eqnarray*}
\\ (2) For $\tilde
r_2(x)$ in (\ref{eqno(3.9)}), under Condition (C4), if $1/5\leq\eta<1$, then
\begin{eqnarray*}
\sqrt{n^{1-\eta}}\Big(\tilde
r_2(x)-r(x)-n^{-2\eta}d(x)\sum\limits_{k=1}^mw_k
\tau_k^2\Big)\stackrel{\cal D}\longrightarrow N\Big(0, {\bf
w}^TA_2{\bf w}\frac{\sigma^2}{f_X(x)}\Big),\
x\in(0,1),\end{eqnarray*} where $d(x)$ is a given function.
 }

{\bf Remark 4.4.} We have the further remarks on the ACR estimators beyond the above comments on conditions (C4) and (C5).
\begin{itemize}
\item[(a)]
{\it Convergence acceleration}. The above result about $\tilde r_1(x)$  shows the importance of bias reduction. As is
known, the kernel estimation is biased, which is the case for all nonparametric smoothers in the literature. To achieve an optimal convergence rate and the asymptotic normality, bandwidth selection must balance between bias and variance terms. Under Condition (C4), the bias $c_n(x)$ of the N-W
estimator has the classical optimal rate $O(n^{-2\eta})$ and  is impossible to be improved through selecting a kernel function because the smoothness assumption is only up to the second order derivative. 
In contract, the
bias  of $\tilde r_1(x)$ is $o(n^{-2\eta})$. This rate-accelerated bias
can then play a very important role for us to get a convergence rate faster
than the classical optimal rate. That is, when the bandwidth  is selected to be
$h=O(n^{-1/5})$, the typical optimal convergence rate of the N-W estimator is
$O(n^{-2/5})$, whereas the ACR estimator $\tilde r_1(x)$
has the rate of $o(n^{-2/5})$. Under Condition (C5), when
the optimal bandwidth
$h=O(n^{-1/9})$ is used, $\tilde r_1(x)$ behaves like the N-W estimator constructed by higher order kernel; both estimators have the same convergence rate of order $O(n^{-4/9}).$
It is worth pointing out that, without use of higher kernel function, the N-W estimator is not possible to have this rate. Thus, under the same conditions on the smoothness of the regression function and kernel function, the estimator $\tilde r_1(x)$ has a faster convergence rate than the classical N-W estimator does. Also it will be shown later that, by the optimal weight, the limiting variance of the ACR estimator $\tilde r_1(x)$ can be smaller than that of the N-W estimator. For $\tilde r_2(x)$, the convergence rate cannot be faster. However, we will verify the estimation efficiency can be promoted as well.
\item[(b)] {\it Weight selection and estimation efficiency}. Invoking the same argument as in
Remark 4.3, we have that the optimal weight vector for the second
estimator $\tilde r_2(x)$ can be expressed  as
$$
{\bf w}_2^*=\left({\bf 1}^TA_2^{-1}{\bf  1}\right)^{-1}A_2^{-1}{\bf
1}.$$ However, $A_1({\bf w})$ for  the first estimator $\tilde
r_1(x)$ depends on the weight vector $\bf w$ as well. Thus,  $\tilde
r_1(x)$ has no a closed form for the corresponding optimal weight
vector. To handle the problem, we approximate $A_1({\bf w})$ by
$$A_1=\left(\frac{ s_ks_j}{\tau_k\tau_j}\int
K\Big(\frac{u}{\tau_k}\Big)K\Big(\frac{u}{\tau_j}\Big)du\right)_{k,j=1}^m,$$
where $s_k=1-\frac{\overline{\tau^2} (\tau_k^2-\overline{\tau^2})}
{\sum\limits_{k=1}^m(\tau_k^2-\overline{\tau^2})^2}$ is free of
the weights vector $\bf w$. A ``sub-optimal"
weight vector for  $\tilde r_1(x)$ with the above $A_1$ is  then $$ {\bf
w}_1^*=\left({\bf 1}^TA_1^{-1}{\bf 1}\right)^{-1}A_1^{-1}{\bf 1}.$$
With the  weights ${\bf
w}_1^*$ and ${\bf
w}_2^*$, $\sqrt{n^{1-\eta}}(\tilde r_1(x)-r(x))$
and $\sqrt{n^{1-\eta}}(\tilde r_2(x)-r(x))$ have the limiting
variances as \begin{equation}\label{(var)}\left({\bf 1}^TA_1^{-1}{\bf
1}\right)^{-1}\frac{\sigma^2}{f_X(x)} \ \mbox{ and } \ \left({\bf
1}^TA_2^{-1}{\bf 1}\right)^{-1}\frac{\sigma^2}{f_X(x)},\end{equation}
respectively. The two limiting variances may be smaller than those of the common kernel estimators. For example, when kernel function is chosen as $K(u)=e^{-\frac{u^2}{2}}/\sqrt{2\pi}$, then
$$A_1=\left(\frac{s_ks_j}{(2\pi)^{1/2}\sqrt{\tau_k^2+\tau_j^2}}\right)_{k,j=1}^m, \ A_2=\left(\frac{1}{(2\pi)^{1/2}\sqrt{\tau_k^2+\tau_j^2}}\right)_{k,j=1}^m.$$ It is known that when the kernel function is chosen as the above,
the limiting variance of the N-W estimator is
$\frac{\sigma^2}{2\sqrt{\pi}f_X(x)}$, which is just a special case of the variances in (\ref{(var)}) with $m=1$ and $\tau_1=1$. Thus, when $\min\{\tau_k;k=1,\cdots,m\}<1<\max\{\tau_k;k=1,\cdots,m\}$ and the above weights are used, the limiting variances of the ACR estimators are smaller than that of the N-W estimator.
\item[(c)] {\it Kernel selection}. As commented above, the ACR estimators can have either faster rate or smaller limiting variance. From the proof, we can see that the estimators  are still the  kernel estimation types. A natural concern is whether the classical N-W estimator could also enjoy this rate-acceleration property through a delicate selection of kernel function. However, when looking into the detail of the proof, we can see that for a single N-W estimator, it is not possible to find such a kernel function, while it does be due to the composition. Therefore, this does show the advantage of the ACR.
%
%
\end{itemize}

We now deal with the asymptotic property of the composite empirical likelihood estimator defined in
(\ref{eqno(block)}). Assume  the following condition:
\begin{itemize}\item[(C6)]
$\sqrt{n}(\bar U(\theta^0,\tau_1),\cdots,\bar U(\theta^0,\tau_m))^T\stackrel{\cal D}\longrightarrow N(0,A_3(\theta^0))$, where $A_3(\theta^0)$ is a positive definite matrix.
\end{itemize}
Clearly, this condition is mild for some common types of dependent data because $\bar U(\theta^0,\tau)$ is actually an average of some functions with zero mean; see, e.g., \citet{Dimitris and Joseph}, and \citet{Lin-Zh}.

{\bf Theorem 4.5} {\it Under Condition (C6), the composite blockwise empirical
likelihood estimator (\ref{eqno(block)}) satisfies
\begin{eqnarray*} \sqrt{n}(\tilde\theta-\theta^0)
\stackrel{\cal D}\longrightarrow N(0,\Delta^{-2}(\theta^0){\bf w}^TA_3(\theta^0){\bf w}).\end{eqnarray*}  }

{\bf Remark 4.5.} Invoking the same arguments as used in Remarks~4.3 and~4.4, the optimal weight vector ${\bf w}^*$ can be designed, and with the optimal weight vector, the ACR estimator is more efficient than the initial estimators; the details are omitted here.

In short, all the theorems above reveal that the ACR method can
 improve original estimators in the sense of either
faster convergence rate or better estimation efficiency. Moreover, by summarizing  all the theorems, we can get the following general conclusions: (1) When $\xi(\tau)$ is free of the model parameter $\theta$, the ARC estimator may be bias-reduced when the original estimator is biased, and consequently the convergence rate could be  faster than the classical one; (2) If  $\xi(\tau)$ depends on $\theta$, the estimation efficiency of the ARC estimator can be improved by choosing proper weights.
These two general conclusions can be  proved theoretically. But complex conditions and expressions are required; the details are thus omitted here.

\setcounter{equation}{0}
\section{ SIMULATION STUDIES}

In this section we examine the finite sample behaviors of the
newly proposed estimators by simulation studies. 
Mean
squared error (for parametric model) and mean integrated squared
error (for nonparametric model) are used to evaluate the efficiency of involved
estimators. We also report the simulation results for
 estimation bias because the initial idea of our method is to
reduce estimation bias.

{\it Experiment 1.} Consider the linear regression  in Example~1. Let $\hat\beta_\tau$ be the common quantile regression estimator
defined in Example 1 and $\tilde \beta$ be the ACE defined
by (\ref{eqno(3.8)}). Here we also consider the composite quantile
regression (CQR) estimator $\hat\beta$ proposed by \cite{r28},
which is constructed by minimizing composite objective function as
\begin{equation}\label{6.1}(\hat\beta^{\tau},\hat b_{\tau_1},\cdots,\hat b_{\tau_m})^T
=\arg\min_{\hspace{-3ex}\beta,b_{\tau_1},\cdots,b_{\tau_m}}\sum_{i=1}^n\sum_{k=1}^m\rho_{\tau_k}
(Y_i-b_{\tau_k}-\beta^{\tau}X_i).\end{equation}  The samples respectively with
size 100, 200 and 400 are generated from the model
$$Y=X^{\tau}\beta+\epsilon,$$ where $\beta=(3,2,1,-1,-2)^T$, the predictors
$X=(X_1,X_2,\cdots,X_5)^T$ follow a multivariate normal distribution
$N(0,\Sigma)$ with $(\Sigma)_{i,j}=0.5^{|i-j|}$ for $1\le i,j\le 5$,
and the error term $\epsilon\sim Gamma(1)$. We choose $\tau=0.5$ to
construct the common quantile regression (QR) estimator
$\hat\beta_\tau$ and select $\tau_k=\frac{k}{10}$ for $k=1,2,\cdots,
9$ to construct both the CQR estimator $\hat\beta$ and the
ACE $\tilde\beta$. 
Empirical bias and mean squared error (MSE) of the three estimators
over 200 replications are reported in Table 1. In
this setting the ACE $\tilde\beta$ is clearly the best one in both bias and variance reduction,
and the QR estimator $\hat\beta_\tau$ is reasonably inferior to the other two competitors.

\begin{table}
\caption{Simulation results in Experiment 1}
\label{tab:1} \vspace{0.3cm} \center
\begin{tabular}{c|ccccccc}
  \hline
  $n$ &   &   & $\hat\beta_1$ & $\hat\beta_2$ & $\hat\beta_3$ & $\hat\beta_4$ & $\hat\beta_5$  \\
  \hline
  \multirow{6}{*}{100} &\multirow{2}{*}{ACE}  & Bias &
  0.0002 & 0.0006 & 0.0011 & $-0.0036$ & 0.0002 \\
  &&MSE& 0.0030 & 0.0035 & 0.0037 & 0.0034 & 0.0025 \\
  \cline{2-8}
    \multirow{6}{*}{} &\multirow{2}{*}{CQR}  & Bias &
  0.0045 & $-0.0024$ & $-0.0030$ & $-0.0019$ & 0.0003 \\
  &&MSE& 0.0059 & 0.0082 & 0.0079 & 0.0085 & 0.0062 \\
  \cline{2-8}
    \multirow{6}{*}{} &\multirow{2}{*}{QR}  & Bias &
  0.0081 & $-0.0066$ & $-0.0057$ & 0.0026 & 0.0004 \\
  &&MSE& 0.0116 & 0.0150 & 0.0167 & 0.0172 & 0.0136 \\
  \cline{2-8}

    \hline
  \multirow{6}{*}{200} &\multirow{2}{*}{ACE}  & Bias &
  0.0007 & $-0.0037$ & 0.0016 & 0.0006 & $-0.0024$ \\
  &&MSE& 0.0012 & 0.0013 & 0.0011 & 0.0012 & 0.0009 \\
  \cline{2-8}
    \multirow{6}{*}{} &\multirow{2}{*}{CQR}  & Bias &
  0.0049 & $-0.0060$ & 0.0021 & 0.0002 & $-0.0024$ \\
  &&MSE& 0.0035 & 0.0036 & 0.0036 & 0.0044 & 0.0029 \\
  \cline{2-8}
    \multirow{6}{*}{} &\multirow{2}{*}{QR}  & Bias &
  0.0033 & $-0.0085$ & 0.0025 & $-0.0002$ & 0.0029 \\
  &&MSE& 0.0074 & 0.0072 & 0.0073 & 0.0077 & 0.0060 \\
  \cline{2-8}

      \hline
    \multirow{6}{*}{400} &\multirow{2}{*}{ACE}  & Bias &
  $-0.0009$ & $-0.0007$ & 0.0020 & $-0.0007$ & 0.0015 \\
  &&MSE& 0.0004 & 0.0006 & 0.0005 & 0.0005 & 0.0004 \\
  \cline{2-8}
    \multirow{6}{*}{} &\multirow{2}{*}{CQR}  & Bias &
  $-0.0063$ & 0.0031 & 0.0023 & $-0.0029$ & 0.0033 \\
  &&MSE& 0.0014 & 0.0016 & 0.0016 & 0.0017 & 0.0015 \\
  \cline{2-8}
    \multirow{6}{*}{} &\multirow{2}{*}{QR}  & Bias &
  $-0.0071$ & $-0.0013$ & 0.0058 & $-0.0053$ & 0.0081 \\
  &&MSE& 0.0035 & 0.0039 & 0.0035 & 0.0043 & 0.0036 \\
  \cline{2-8}

  \hline

\end{tabular}
\end{table}

{\it Experiment 2.} For the nonparametric regression
$$Y_i=r(X_i)+\epsilon_i, i=1,\cdots,n,$$ the common local constant (LC)
estimator (kernel estimator) is defined as
\begin{equation}\label{6.2}\hat r_h(x)=\frac{\sum\limits_{i=1}^nY_iK(\frac{X_i-x}{h})}
{\sum\limits_{i=1}^nK(\frac{X_i-x}{h})}.\end{equation} As a comparison, we here
consider a composite objective function method, which is defined by following way:
for $h_k=\tau_kn^{-\eta}, k=1,\cdots,m$, define a composite
local constant (CLC) estimator as
$$\hat r(x)=\arg\min_{a}\sum_{i=1}^n\sum_{k=1}^m(Y_i-a)^2K
\Big(\frac{X_i-x}{h_k}\Big).$$ This estimator has the following closed
representation:
\begin{equation}\label{6.3}\hat r(x)=\frac{\sum\limits_{i=1}^n\sum\limits_{k=1}^mY_iK
\Big(\frac{X_i-x}{h_k}\Big)}{\sum\limits_{i=1}^n\sum\limits_{k=1}^mK
\Big(\frac{X_i-x}{h_k}\Big)}.\end{equation} Thus such an estimator can be regarded
as an indirect composition of the LC estimators (\ref{6.2}) with different bandwidths. Now we compare the ACE
estimators defined by (\ref{eqno(3.9)}) with the LC estimator and the CLC estimator mentioned above
via simulation studies.
Consider the regression function
$r(X)=\sin(2\pi X)$, $X\sim U(0,1)$, $\epsilon\sim N(0,0.5^2)$ with the
sample size $n= 100$, 200 and 400 respectively. In this experiment,
the Epanechnikov kernel $K(u)=0.75(1-u^2)\bold{1}_{|u|\le 1}$ is
employed, and for simplicity the equal weights are used in the ACE. In the local constant estimation procedure, the bandwidth
$h$ is chosen by two-fold cross-validation. Then $\tau_k$'s are
chosen so that $h_k$'s are around $h$. Simulation results are tabulated
in Table 2, in which MISE denotes the empirical mean integrated
squared errors through 200 replications. By comparing  MISEs of
the three estimators, we see that the ACE behaves the best
among the three estimators even the optimal weights are not
employed. Meanwhile, we notice that the CLC estimator $\hat r(x)$
given in (\ref{6.3}) is the worst one.
The above two findings indicate that the ACE is an efficient
composite estimator, whereas the competitor such as the CLC through a composite objective function is not always
efficient.

\begin{table}
\caption{MISE for nonparametric estimators in Experiment 2}
\label{tab:2} \vspace{0.3cm} \center
\begin{tabular}{c|ccc}
  \hline
    & n=100 & n=200 & n=400 \\
    \hline
  LC  & 0.0228 & 0.0136 & 0.0075 \\
  CLC & 0.0262 & 0.0176 & 0.0121 \\
  ACE & 0.0206 & 0.0121 & 0.0068\\
  \hline
\end{tabular}
\end{table}

{\it Experiment 3.}
Consider the following  linear regression model:
$$Y_i=\theta X_i+\varepsilon_i, \ i=1,\cdots,n,$$ where $\theta$ is a scalar parameter. In this model the errors $\varepsilon_i, i=1,\cdots,n$, are dependent, satisfying
$$\varepsilon_i=a\varepsilon_{i-1}+\epsilon_i, i=2,\cdots,n, \varepsilon_1=\epsilon_1,$$ where $0<|a|<1$ and $\epsilon_i, i=1,\cdots,n$, are independent and identically distributed from $N(0,1)$. In this case, an unbiased estimating function is chosen to be  $u(\theta)=X_i(Y_i-\theta X_i)$. Thus the corresponding blockwise estimating function can be expressed as
$$U_i(\theta)=\frac{1}{M}\sum_{k=1}^{M}X_{(i-1)L_\tau+k}(Y_{(i-1)L_\tau+k}-\theta X_{(i-1)L_\tau+k}).$$

We first consider method 1: the blockwise empirical Euclidean likelihood defined in Subsection 3.3, by which the blockwise empirical Euclidean log-likelihood ratio has the following closed representation:
\begin{equation}\label{likelihood} l_\tau(\theta)=-\frac{Q_\tau}{2 S}\bar U^2(\theta),\end{equation} where $S(\theta)=\frac{1}{Q_\tau}\sum_{i=1}^{Q_\tau}(U_i(\theta)-\bar U(\theta))^2$ and $\bar U(\theta)=\frac{1}{Q_\tau}\sum_{i=1}^{Q_\tau}U_i(\theta)$.
In the simulation, $\theta$
is chosen  to be $\theta=5$, the window-width is  $M=[n^{1/3}]$ (i.e., $c=1/3$), $\tau$ is determined by the CV and the different values for the ACE are taken around $\tau$. We now compare the ACE defined by (\ref{eqno(block)}) and the blockwise empirical likelihood estimation (BELE) obtained by minimizing $l_\tau(\theta)$ in (\ref{likelihood}).
Simulation results tabulated
in Table~3 are obtained through 200 replications. We can see that the ACE behaves slightly better than the BELE does in the sense that the bias and MSE of the ACE are slightly but uniformly smaller than those of the BELE.

To further examine the behaviour of our method, now we consider method 2, an approximate method, as follows. It is known that $S$ in (\ref{likelihood}) is a consistent estimator of the variance of the error. If $S$ is ignored, the blockwise empirical Euclidean log-likelihood ratio has the following approximate representation:
\begin{equation}\label{likelihood2} l_\tau(\theta)\propto -\frac{Q_\tau}{2 }\bar U^2(\theta).\end{equation} In the simulation, $\theta$
is chosen  to be $\theta=2.5$, the window-width is $M=[n^{1/2}]$ (i.e., $c=1/2$), the other conditions are designed as in method 1. We now compare the ACE defined by (\ref{eqno(block)}) and the approximate BELE obtained by minimizing $l_\tau(\theta)$ in (\ref{likelihood2}). The following Table~4  reports the simulation results about bias and MSE for different combinations of $X_i\sim N(0,1)$, $X_i\sim N(0.3,1)$, $a=0.1,0.3,-0.3$ and $n=100,200,300$, respectively. By comparing Table~3 and Table~4, we see that when $S$ is removed from the likelihood ratio, the approximate BELE runs into problems but the ACE still works well. More precisely, we have the following findings: (1)
When $X_i\sim N(0.3,1)$, the behavior of the BELE is relatively stable. The ACE works better than the BELE in the sense that both the bias and the MSE of the new estimator are smaller than those obtained by the BELE; (2) When $X_i\sim N(0,1)$, the MSE of the BELE is quite large showing that the BELE is very unstable. In contrast, the ACE still works very well with much smaller bias and the MSE.

These simulation results and the definition in (\ref{eqno(block)}) show that when the algorithm for obtaining  the initial estimators $\hat\theta_{\tau_k}$ is not stable, the ACE can efficiently improve the performance.  
Thus the ACE is an efficient composite method specially for the case when the original estimator is unstable.

\begin{table}
\caption{Simulation results of method 1 in Experiment 3}
\label{tab:3} \vspace{0.3cm} \center
\begin{tabular}{c|c|c|cc|cccc}
  \hline
 \multirow{2}{*}{$X$} &\multirow{2}{*}{$n$} &  \multirow{2}{*}{$a$}  & BELE  & & ACE  &  \\
  & &  &  Bias& MSE  &  Bias & MSE  \\
  \hline
  &\multirow{3}{*}{100} &0.1  & $-0.0081$ & 0.0121 & $0.0055$ & 0.0111 \\

 & & 0.5 & 0.0017 & 0.0145 &   $-0.0013$ & 0.0133\\

 & &0.9 & $-0.0018$& 0.0696 &  $-0.0006$ & 0.0661 \\

  \cline{2-7}
 \multirow{3}{*}{$N(0,1)$}& \multirow{3}{*}{200} &0.1  & $-0.0110$ & 0.0048 & $-0.0106$ & 0.0046 \\

 & & 0.5 &    $-0.0070$ & 0.0072 &  $-0.0051$& 0.0070 \\

  &  &0.9 &  $-0.0055$ & 0.0331 &   $-0.0053$ & 0.0317 \\

  \cline{2-7}
 & \multirow{3}{*}{400} &0.1  & 0.0057& 0.0033 &  $-0.0054$ &0.0031 \\

 & & 0.5 & $-0.0043$ & 0.0034 & $-0.0038$ & 0.0032 \\
  &  &0.9 &  0.0034 & 0.0136 &  0.0015 & 0.0130 \\
  \hline

\end{tabular}
\end{table}

\begin{table}
\caption{Simulation results of method 2 in Experiment 3}
\label{tab:3} \vspace{0.3cm} \center
\begin{tabular}{c|c|c|cc|cccc}
  \hline
 \multirow{2}{*}{$X$} &\multirow{2}{*}{$n$} &  \multirow{2}{*}{$a$}  & BELE  & & ACE  &  \\
  & &  &  Bias& MSE  &  Bias & MSE  \\
  \hline
  &\multirow{3}{*}{100} &0.1  & 0.0222 & 0.4264   &0.0020 & 0.0204 \\

 & & 0.3 & $-0.0408$ & 0.2997 &   $-0.0052$ & 0.0244\\

 & &$-0.3$ & $-0.0644$ & 2.9066 &  $-0.0088$ & 0.0279 \\

  \cline{2-7}
 \multirow{3}{*}{$N(0.3,1)$}& \multirow{3}{*}{200} &0.1  & $-0.0089$ & 0.0398 & 0.0076 & 0.0090 \\

 & & 0.3 &    0.0108 & 0.1749 &  0.0075& 0.0127 \\

  &  &$-0.3$ &  $-0.0117$ & 0.0272 &   0.0050 & 0.0101 \\

  \cline{2-7}
 & \multirow{3}{*}{300} &0.1  & 0.0028& 0.0209 &  0.0020 &0.0060 \\

 & & 0.3 & 0.0057 & 0.0310 & 0.0026 & 0.0080 \\
  &  &$-0.3$ &  0.0041 & 0.0159 &  0.0020 & 0.0064 \\
  \hline
  &\multirow{3}{*}{100} &0.1  & $-0.3241$ & $17.9075$   &$-0.0110$ & 0.0253 \\

 & & 0.3 & $0.0449$ & 17.3460 &   0.0037 & 0.0271\\

 & &$-0.3$ & $-0.2882$ & 16.1170 &  $-0.0012$ & 0.0235 \\

  \cline{2-7}
 \multirow{3}{*}{$N(0,1)$}& \multirow{3}{*}{200} &0.1  & 0.1270 & 30.6297 &  0.0153 & 0.0124 \\

 & & 0.3 &    0.1569 & 104.3950 &  $-1.9725\mbox{e}-5$& 0.0142 \\

  &  &$-0.3$ &  0.2147 & 10.3675 &   $-0.0021$ & 0.0120 \\

  \cline{2-7}
 & \multirow{3}{*}{300} &0.1  &  0.3470 & 28.7330 &  0.0047 &0.0065 \\

 & & 0.3 & 0.1275 & 11.3672 & $-0.0030$ & 0.0063 \\
  &  &$-0.3$ &  0.0563 & 13.9492 &  $-0.0061$ & 0.0057 \\
  \hline

\end{tabular}
\end{table}

\newpage





\begin{thebibliography}{1}



\bibitem[Bahadur(1966)]{Bahadur} Bahadur, R. R. (1966), ``A note on quantiles in Large Samples'', {\it Ann.
Math. Statist.}, {\bf 37}, 577-580.

\bibitem[Bhattacharya and Gangopadhyay(1990)]{Bhattacharya-G} Bhattacharya, P. K. and Gangopadhyay, A.
(1990), ``Kernel and nearest neighbor estimation of a conditional
quantile'', {\it Ann. Statist.}, {\bf 18}, 1400-1415.
%
%
\bibitem[Chaudhuri(1991)]{Chaudhuri} Chaudhuri, P. (1991), ``Nonparametric estimates of regression
quantiles and their local Bahadur representation'', {\it Ann.
Statist.}, {\bf 19}, 760-777.

\bibitem[Dimitris and Joseph(1992)]{Dimitris and Joseph} Dimitris, N., Joseph, P., (1992), ``A general resampling scheme for triangular arrays of $\alpha$-mixing random variables with application to the
problem of spectral density estimation'', {\it Ann. Statist}., {\bf
20}, 1985-2007.

%
%
\bibitem[Fan and Wang(2011)]{Fan-W} Fan, J. and Wang, W. (2011). Penalized composite
quasi-likelihood for ultrahigh dimensional variable selection. {\it
J. R. Statist. Soc.} B, {\bf 73}, 325-349.
%
%
%
\bibitem[Gray and Schucany(1972)]{Gray-Sch}  Gray, H., Schucany, W. R. (1972), {\it The generalized jackknife
statistic.}, New York, M. Dekker.
%
\bibitem[Hansen(2007)]{Hansen} Hansen, B. E. (2007), ``Least squares model averaging'', {\it
Econometrica}, {\bf 75}, 1175-1189.
%
\bibitem[Hoeting {\it et al.}(1999)]{Hoeting} Hoeting, J. A., Madigan, D., Raftery, A. E. and
Volinsky, C. T. (1999), ``Bayesian model averaging: A tutorial'',
{\it Statistical Science}, {\bf 14}, 382-417. Correction, {\bf 15},
193-195.
%
\bibitem[Hong(2003)]{Hong} Hong, S. Y. (2003), ``Bahadur representation and its
applications for local polynomial estimation in nonparametric
$M$-regression'', {\it Nonparametric Statistics}, {\bf 15}, 237-251.
%

\bibitem[Huang, Ma and Zhang(2008)]{Huang-M-Zh} Huang, J, Ma, S. and Zhang, C. (2008), ``Adaptive lasso for sparse high-dimensional regression models'', {\it Statistica Sinica}, {\bf 18}, 1603-1618.

%
%
%
%
\bibitem[Kai, Li and Zou(2010)]{Kai-L-Z0} Kai, B, Li, R. and Zou, H. (2010), `` Local composite quantile regression
smoothing: an efficient and safe alterative to local polynomial
regression'', {\it J. R. Statist. Soc.} B {\bf 72}, 49-69.
%
\bibitem[Kai, Li and Zou(2011)]{Kai-L-Z1} | (2011), ``
New efficient estimation and variable selection methods for
semiparametric varying-coefficient partially linear models'', {\it
Ann. Statist.}, {\bf 39}, 305-332.


%
\bibitem[Kiefer(1967)]{Kiefer} Kiefer, J. (1967), ``On Bahadur"s representation of sample quantiles'',
{\it Ann. Math. Statist.}, {\bf 38}, 1323-1342.
%
%
\bibitem[Kitamura(1997)]{Kitamura} Kitamura, Y. (1997), ``Empirical likelihood methods with weakly dependent processes'',
{\it Ann. Statist.}, {\bf 25}, 2084¨C2102.

\bibitem[Koenker(1984)]{Koenker-1984} Koenker, R. (1984), ``A note on L-estimates for linear models'', {\it Statistics and
Probability Letters}, {\bf 2}, 323-325.

%
\bibitem[Koenker(2005)]{Koenker} | (2005), {\it Quantile Regression}. Cambridge Univ. Press.
%
%
\bibitem[Lin and Zhang(2001)]{Lin-Zh} Lin, L. and Zhang, R. C. (2001), ``Blockwise empirical Euclidean likelihood for weakly
dependent processes'', {\it Statistics and Probability Letters},
{\bf 53},143-52.

\bibitem[Miller(1974)]{Miller} Miller, R. G. (1974), ``The jackknife--a review'', {\it Biometrika}, {\bf 61}, 1-15.
%
%
\bibitem[Pollard(1984)]{Pollard} Pollard, D. (1984), {\it Convergence of stochastic processes}.
New York, Berlin Heidelberg, Tokyo.


\bibitem[Quenouille(1949)]{Quenouille1} Quenouille, M. H. (1949), ``Approximate tests of correlation in time-series'', {\it J. R. Statist. Soc.} B, {\bf 11}, 68-84.

\bibitem[Quenouille(1956)]{Quenouille2} | (1956), ``Notes on bias in estimation'', {\it Biometrika}, {\bf 43}, 353-360.


%
%
\bibitem[Serfling(1980)]{Serfling} Serfling, R. J. (1980), {\it Approximation Theorem of Mathematical
Statistics}. John Wiley \& Sons, Inc.
%
%
\bibitem[Shao(1991)]{Shao} Shao, J. (1991), ``Second-order differentiability and jackknife'', {\it
Statist. Sinica}, {\bf 1}, 185-202.
%
\bibitem[Sun, Gai and Lin(2013)]{Sun-L} Sun, J., Gai Y. J. and Lin, L. (2013), ``Weighted local linear composite quantile
estimation for the case of general error distributions'', {\it
Journal of Statistical Planning and Inference}, {\bf 143},
1049-1063.

%
%
\bibitem[Tibshirani(1996)]{Tibshirani} Tibshirani, R. (1996), ``Regression shrinkage and selection via the Lasso'',
{\it J. R. Statist. Soc.}, B, {\bf 58}, 267-288.

\bibitem[Tuky(1958)]{Tuky} Tuky, J. W. (1958), ``Bias and condidence in not-quite large smaples (Abastract)'', {\it Ann. Math. Statist.}, {\bf 29}, 614.
%
\bibitem[Varin, Reid and Firth(2011)]{Varin-R-F} Varin, C., Reid, N. and Firth, D. (2011), ``An overview of composite likelihood methods'', {\it Statist. Sinica}, {\bf 21}, 5-42.
%
\bibitem[van der Vaart(1998)]{Van der} van der Vaart, A. W. (1998). {\it Asymptotic Statistics}. Cambridge University
Press.
%
\bibitem[Wang, Zhang and Zou(2010)]{Wang-Z-Z} Wang, A. T. K., Zhang, X. and Zou, G. (2010), ``Least squares
model averaging by mallows criterion'', {\it Journal of
Econometrics}, {\bf 156}, 277-283.
%
\bibitem[Wainwright(2009)]{Wainwright} Wainwright, M. J. (2009), ``Sharp threshold for high-dimensional and noisy sparsity recovery using
$\ell_1$-constrained quadratic programming (Lasso)'', {\it IEEE
Transactions on Information Theory}, {\bf 55}, 2183-2202.
%
%
%
%
%
\bibitem[Zou and Yuan(2008)]{Zou-Y} Zou, H. and Yuan, M. (2008), ``Composite quantile regression and
the oracle model selection theory'', {\it Ann. Statist.}, {\bf 36},
1108-1126.

\bibitem[Zhu(1993)]{Zhu} Zhu, L.-X. (1993), ``Convergence rates of empirical processes indexed by classes of functions and their applications'', {\it J. Syst. Sci. Math. Sci.}, {\bf 13} 33 - 41.

\end{thebibliography}
\end{document}